\documentstyle[epsf,12pt]{article}

\makeatletter

 \@addtoreset{equation}{section}
\makeatother

\begin{document}
\topmargin 0pt
\oddsidemargin 5mm

\newcommand {\beq}{\begin{eqnarray}}
\newcommand {\eeq}{\end{eqnarray}}
\newcommand {\non}{\nonumber\\}
\newcommand {\eq}[1]{\label {eq.#1}}
\newcommand {\defeq}{\stackrel{\rm def}{=}}
\newcommand {\gto}{\stackrel{g}{\to}}
\newcommand {\hto}{\stackrel{h}{\to}}
\newcommand {\1}[1]{\frac{1}{#1}}
\newcommand {\2}[1]{\frac{i}{#1}}
\newcommand {\th}{\theta}
\newcommand {\thb}{\bar{\theta}}
\newcommand {\ps}{\psi}
\newcommand {\psb}{\bar{\psi}}
\newcommand {\ph}{\varphi}
\newcommand {\phs}[1]{\varphi^{*#1}}
\newcommand {\sig}{\sigma}
\newcommand {\sigb}{\bar{\sigma}}
\newcommand {\Ph}{\Phi}
\newcommand {\Phd}{\Phi^{\dagger}}
\newcommand {\Sig}{\Sigma}
\newcommand {\Phm}{{\mit\Phi}}
\newcommand {\eps}{\varepsilon}
\newcommand {\del}{\partial}
\newcommand {\dagg}{^{\dagger}}
\newcommand {\pri}{^{\prime}}
\newcommand {\prip}{^{\prime\prime}}
\newcommand {\pripp}{^{\prime\prime\prime}}
\newcommand {\prippp}{^{\prime\prime\prime\prime}}
\newcommand {\pripppp}{^{\prime\prime\prime\prime\prime}}
\newcommand {\delb}{\bar{\partial}}
\newcommand {\zb}{\bar{z}}
\newcommand {\mub}{\bar{\mu}}
\newcommand {\nub}{\bar{\nu}}
\newcommand {\lam}{\lambda}
\newcommand {\lamb}{\bar{\lambda}}
\newcommand {\kap}{\kappa}
\newcommand {\kapb}{\bar{\kappa}}
\newcommand {\xib}{\bar{\xi}}
\newcommand {\ep}{\epsilon}
\newcommand {\epb}{\bar{\epsilon}}
\newcommand {\Ga}{\Gamma}
\newcommand {\rhob}{\bar{\rho}}
\newcommand {\etab}{\bar{\eta}}
\newcommand {\chib}{\bar{\chi}}
\newcommand {\tht}{\tilde{\th}}
\newcommand {\zbasis}[1]{\del/\del z^{#1}}
\newcommand {\zbbasis}[1]{\del/\del \bar{z}^{#1}}
\newcommand {\vecv}{\vec{v}^{\, \prime}}
\newcommand {\vecvd}{\vec{v}^{\, \prime \dagger}}
\newcommand {\vecvs}{\vec{v}^{\, \prime *}}
\newcommand {\alpht}{\tilde{\alpha}}
\newcommand {\xipd}{\xi^{\prime\dagger}}
\newcommand {\pris}{^{\prime *}}
\newcommand {\prid}{^{\prime \dagger}}
\newcommand {\Jto}{\stackrel{J}{\to}}
\newcommand {\vprid}{v^{\prime 2}}
\newcommand {\vpriq}{v^{\prime 4}}
\newcommand {\vt}{\tilde{v}}
\newcommand {\vecvt}{\vec{\tilde{v}}}
\newcommand {\vecpht}{\vec{\tilde{\phi}}}
\newcommand {\pht}{\tilde{\phi}}
\newcommand {\goto}{\stackrel{g_0}{\to}}
\newcommand {\tr}{{\rm tr}\,}
\newcommand {\GC}{G^{\bf C}}
\newcommand {\HC}{H^{\bf C}}
\newcommand{\vs}[1]{\vspace{#1 mm}}
\newcommand{\hs}[1]{\hspace{#1 mm}}

\setcounter{page}{0}

\begin{titlepage}

\begin{flushright}
OU-HET 334\\
hep-th/9911225\\
November 1999
\end{flushright}
\bigskip

\begin{center}
{\LARGE\bf
Quantum Equivalence of Auxiliary Field Methods 
in Supersymmetric Theories
}
\vs{10}

\bigskip
{\renewcommand{\thefootnote}{\fnsymbol{footnote}}
{\large\bf Kiyoshi Higashijima\footnote{
     E-mail: higashij@phys.sci.osaka-u.ac.jp}
 and Muneto Nitta\footnote{
E-mail: nitta@het.phys.sci.osaka-u.ac.jp
}
}
}

\setcounter{footnote}{0}
\bigskip

{\small \it
Department of Physics,
Graduate School of Science, Osaka University,\\
Toyonaka, Osaka 560-0043, Japan
}
\end{center}
\bigskip

\begin{abstract}
Quantum corrections to Legendre transformations 
are shown to cancel to all orders in supersymmetric 
theories in path integral formalism. 
Using this result, lagrangians for auxiliary fields 
are generalized to non-quadratic forms. 
In supersymmetric effective nonlinear lagrangians, 
the arbitrariness due to the existence of quasi 
Nambu-Goldstone bosons is shown to disappear 
when local auxiliary gauge fields are introduced.
\end{abstract}

\end{titlepage}

\newpage
\section{Introduction}
The Legendre transformation plays an important role both 
in classical and quantum physics. It appears when we 
change independent variables in classical mechanics. 
In quantum physics, it is associated with the Fourier 
transformation. For example, 
let us consider the path integral 
\beq
  I = \int [d\sig] \exp\left[i \int d^4x   
            \;(\sig^i \Phi_i - W(\sig^1,\cdots,\sigma^n)) \right],
\eeq
where $\sig^i(x)$ and $\Phi_i(x)$ ($i=1,\cdots,n$) 
are real scalar fields and 
$W(\sig^1,\cdots,\sig^n)$ is an arbitrary function.  
In order to evaluate this integral, we expand the integrand 
around the stationary path $\hat{\sig}$ 
defined by 
\beq
 {\del \over \del \sig^i} 
 (\sig^j \Phi_j - W (\sig^1\cdots\sig^n)) |_{\sig^i =\hat\sig^{i}} 
 = \Phi_i - W_i (\hat\sig^{1},\cdots,\hat\sig^{n}) = 0  
     \label{nonSUSY_st.},
\eeq
where the subscript $i$ denotes 
differentiation with respect to $\sig^i$.  
Then the integral can be performed by the saddle point method:  
\beq
 &&I = \exp\left[i \int d^4x\; 
   \bigg(\hat\sig^{i} \Phi_i 
      - W(\hat\sig^{1},\cdots,\hat\sig^{n})\bigg)\right] \non 
 &&\times \int [d\sig] \exp \left[ -i \int d^4x\;  
  \sum_{ij}{1\over 2}
  {\del^2 W(\sig^1,\cdots,\sig^n) \over \del \sig^i \del\sig^j}
   |_{\sig^i = \hat\sig^{i}}
  (\sig - \hat\sig)^i (\sig -\hat\sig)^j + \cdots \right]. \hspace{1cm} 
\eeq
Here the first factor is the classical contribution and 
the second factor consists of quantum corrections.
Linear terms are absent because of Eq.~(\ref{nonSUSY_st.}). 
The quadratic terms provide a factor of  
$(\det {\bf W}(\hat\sig))^{-\1{2}}$ after the Gaussian integration,  
with $({\bf W})_{ij} = {\del^2 W \over \del \sig^i \del \sig^j}$ 
being the Hesse matrix. 
The remaining terms contribute to higher order quantum corrections. 
Thus, if we define $U(\Phi_i)$ by 
\beq
  I = \exp\left[ iU(\Phi_i)\right],
\eeq
then the classical expression
\beq
  U(\Phi_i)=\hat\sig^{j} \Phi_j - W (\hat\sig^{1},\cdots,\hat\sig^{n})
\eeq
is just the lowest order approximation. Unless $W(\sigma)$ 
is a quadratic form, there are many higher order quantum 
corrections in general.
In supersymmetric theories, however, we show this 
{\it classical expression is exact to all orders} 
because of remarkable cancellations of quantum 
corrections due to the supersymmetry.

This quantum Legendre transform may have many applications.
For example, it can be applied to studies of 
supersymmetric nonlinear sigma models~\cite{Zu}. 
They can be defined from linear sigma models 
by introducing auxiliary fields 
as lagrange multipliers. 
For example, consider a lagrangian 
${\cal L} = \int d^4 \th K_0(\phi,\phi\dagg) + 
(\int d^2 \th P(\phi) + {\rm c.c.} )$, 
with the canonical K\"{a}hler potential 
$K_0 = \phi\dagg\phi$ and the superpotential 
$P = \phi_0 g(\phi)$.
Here, $\phi_0$ is an auxiliary field 
without kinetic terms.
The path integral over $\phi_0$ gives a F-term constraint, 
$g(\phi) = 0$. 
By solving this equation and 
substituting the solution into $K_0$, 
we obtain the K\"{a}hler potential 
for a nonlinear sigma model. 
If we consider the above ``linear model'' 
as an effective theory, we have to use 
the more general K\"{a}hler potential 
$K_0 = f(\phi\dagg\phi)$, 
where $f$ is an arbitrary function~\cite{BKMU}.  
Then the resulting K\"{a}hler potential 
is $K = f(\phi\dagg\phi)|_{g(\phi)=0}$~\cite{Ni1,Ni2} 
after integrating out $\phi_0$.

In order to obtain a supersymmetric nonlinear sigma model 
on a {\it compact} manifold, we have to consider a gauged 
linear sigma model by introducing a vector superfield $V$ 
as an auxiliary field:\footnote{
Although, we discuss the Abelian 
model for simplicity in the Introduction, 
extension to the non-Abelian case is discussed below.} 
$K_0= e^V\phi\dagg\phi - cV$, 
where $c$ is called a Fayet-Iliopoulos parameter. 
By solving the classical equation of motion for $V$, 
$e^V \phi\dagg\phi - c = 0$, 
and substituting back into $K_0$, 
we obtain a K\"{a}hler potential 
$K= c\log(\phi\dagg\phi) + {\rm const}$, 
which is no longer linear~\cite{Ao,Ku}. 
This procedure can be considered as 
a K\"{a}hler quotient method~\cite{HKLR},  
and it is used to construct low energy effective theories of 
supersymmetric gauge theories~\cite{gauge}.
As an application of our exact quantum Legendre transform, 
we will show that {\it use of the classical equation of motion 
for $V$ can be justified} in the path integral formalism.
Moreover, as stated above, 
the K\"{a}hler potential can be generalized to 
$K = f(e^V \phi\dagg \phi) -cV$ 
with an arbitrary function $f$. 
One of the main results of this paper is that we can show that 
this {\it arbitrariness disappears} 
after integrating out the vector superfield. 

As an example, by combining these two auxiliary field methods, 
we construct a nonlinear sigma model, whose target space is  
the compact homogeneous K\"{a}hler manifold 
$SO(N)/SO(N-2)\times U(1)$, from a linear model.
The construction of 
the hermitian symmetric spaces from linear models 
is discussed in a separate paper~\cite{HN}. 

This paper is organized as follows. In section 1 we prove 
a theorem on the quantum equivalence of auxiliary field 
methods in supersymmetric theories. 
In section 2 we apply 
this theorem to supersymmetric nonlinear sigma models.

\section{Quantum Legendre transformation}
Before proving the absence of quantum corrections, 
we summarize useful formulas for our proof. 
Vector superfields (real superfields) satisfy $V\dagg = V$, 
and can be written in component fields as~\cite{WB} 
\beq
 V(x,\th,\thb)
 &=& C(x) + i\th \chi(x) - i \thb \chib(x)\non
 &&+{i \over 2} \th\th [M(x) + iN(x)]
 -{i \over 2} \thb\thb [M(x) - iN(x)]\non
 &&- \th \sigma^{\mu} \thb v_{\mu}(x) 
 + i \th\th\thb \lamb(x)
 - i \thb\thb\th \lam(x)
 + \1{2} \th\th\thb\thb D(x) .
\eeq
The D-term of the product of two vector superfields is 
\beq
 V^i V^j |_{\th\th\thb\thb}
 &=& \1{2} (C^i D^j + C^j D^i)
  -\1{2} (\chi^i\lam^j + \chib^i\lamb^j) 
  -\1{2} (\chi^j\lam^i + \chib^j\lamb^i) \non
 && +\1{2} (M^iM^j + N^iN^j - v^i \cdot v^j)  .\label{product}
\eeq
Next an arbitrary function of $n$ vector superfields, 
$f(V^1,\cdots,V^n)$, can be expanded in a Taylor series 
around $C = (C^1,\cdots,C^n)$ as  
\beq
 &&f(V^1, \cdots, V^n) \non
 && = f(C) + i f_i(C)(\th\chi^i - \thb \chib^i) \non
 &&+ \2{2} \th\th \left[f_i(C)(M^i+iN^i) 
           - \2{2}f_{ij}(C)\chi^i\chi^j \right] \non
 &&- \2{2} \thb\thb 
    \left[ f_i(C)(M^i-iN^i) 
            + \2{2} f_{ij}(C)\chib^i\chib^j \right] \non
 &&- \th \sig^{\mu} \thb 
  \left[f_i(C){v^i}_{\mu} 
       + \1{2}f_{ij}(C)
         (\chi^{(i} \sigma_{\mu} \chib^{j)}) \right]\non
 &&+ i \th\th\thb
   \left[ f_i(C)\lamb^i - \2{2}f_{ij}(C)\chib^{(i}(M+iN)^{j)} 
         -\1{2}f_{ij}(C) \sigb^{\mu} \chi^{(i} {v^{j)}}_{\mu} 
         -\1{6}f_{ijk}(C) 
              \chib^{(i}\chi^j\chi^{k)} \right]\non
 &&-i \thb\thb\th
   \left[f_i(C)\lam^i + \2{2}f_{ij}(C)\chi^{(i}(M-iN)^{j)} 
         +\1{2}f_{ij}(C) \sig^{\mu} \chib^{(i} {v^{j)}}_{\mu} 
         -\1{6}f_{ijk}(C)\chi^{(i}\chib^j\chib^{k)} \right]\non
 &&+ \1{2} \th\th\thb\thb
   \left[f_i(C) D^i 
       - f_{ij}(C)(\chi^{(i}\lam^{j)} + \chib^{(i}\lamb^{j)}) 
       + \1{2} f_{ij}(C) (M^iM^j + N^iN^j -  v^i\cdot v^j ) 
          \right.\non
  &&\hspace{1cm} \left.
        -\1{4} f_{ijk}(C) 
       \{ i(\chi^{(i}\chi^j - \chib^{(i}\chib^j) M^{k)} 
        +  (\chi^{(i}\chi^j + \chib^{(i}\chib^j) N^{k)}
        +  2 {v^{(i}}_{\mu}(\chi^j\sig^{\mu}\chib^{k)}) \} 
    \right.\non
  &&\hspace{1cm} \left.
   + \1{8}f_{ijkl}(C) \chi^{(i}\chi^j\chib^k\chib^{l)} 
                      \right], \label{f(V)}
\eeq
where we have used the notation 
\beq
 A^{(i_1\cdots i_n)} 
  = \1{n !}(A^{i_1\cdots i_n} + ({\rm symmetrization})),
\eeq
and the subscripts $i,j,\cdots$ denote 
differentiation with respect to $C^i,C^j,\cdots$.

\bigskip
Now we are ready to state the main result of this 
section regarding the absence of quantum corrections to the 
Legendre transform. 

{\bf Theorem 1.} (The quantum Legendre transformation.)\\
Let $\sig^i(x,\theta,\bar{\theta})$ and 
$\Phi_i(x,\theta,\bar{\theta})$ ($i=1,\cdots,n$) 
be vector superfields and $W$ be a function of $\sig^i$. 
Then,  
\beq
 \int [d\sig] \exp\left[i \int d^4x d^4\th  
    \;(\sig^i \Phi_i - W(\sig^1,\cdots,\sig^n)) \right]
 = \exp\left[i \int d^4x d^4\th  \;U(\Phi) \right] .
\eeq
Here $U(\Phi)$ is defined by 
\beq
 U(\Phi) = \hat{\sig}^{i} (\Phi) \Phi_i 
  - W(\hat{\sig}^{1} (\Phi),\cdots,\hat{\sig}^{n} (\Phi)) ,
\eeq
where $\hat{\sig}$ is a solution of the stationary equation 
\beq
 {\del \over \del \sig^i} 
  (\sig^j \Phi_j - W (\sig^1,\cdots,\sig^n)) |_{\sig^i =\hat\sig^{i}} 
 = \Phi_i - W_i (\hat\sig^{1},\cdots,\hat\sig^{n}) = 0 
     \label{sigb} .
\eeq
(Proof)
To show this, we calculate the path integral of $\sigma$ 
on the left-hand side explicitly, and then 
compare with the right-hand side. 
By using Eqs.~(\ref{product}) and (\ref{f(V)}), 
the integrand of the left-hand side can be written as 
the exponential of 
\beq
 &&\hspace{-0.5cm} 
 (\sig^i \Phi_i - W(\sig^1,\cdots,\sig^n))|_{\th\th\thb\thb} \non
 &=& \1{2} D_{\sig}^i (C_{\Phi i} - W_i(C_{\sig}) )
  -\1{2} \lam_{\sig}^i 
     (\chi_{\Phi i} - W_{ij}(C_{\sig})\chi_{\sig}^j)
  -\1{2} \lamb_{\sig}^i 
     (\chib_{\Phi i} - W_{ij}(C_{\sig})\chib_{\sig}^j)\non
 && +\1{2} C_{\sig}^i D_{\Phi i}
    -\1{2} (\chi_{\sig}^i \lam_{\Phi i} 
          + \chib_{\sig}^i \lamb_{\Phi i})
    +\1{2} (M_{\sig}^iM_{\Phi i} + N_{\sig}^iN_{\Phi i} 
           -v_{\sig}^i \cdot v_{\Phi i}) \non 
 && -\1{4} W_{ij}(C_{\sig}) 
     ( M_{\sig}^iM_{\sig}^j + N_{\sig}^i N_{\sig}^j 
     - v_{\sig}^i \cdot v_{\sig}^j )\non
 && +\1{8} W_{ijk}(C_{\sig}) 
     \left[i(\chi_{\sig}^{(i}\chi_{\sig}^j 
            -\chib_{\sig}^{(i}\chib_{\sig}^j )M_{\sig}^{k)}  
          + (\chi_{\sig}^{(i}\chi_{\sig}^j  
            +\chib_{\sig}^{(i}\chib_{\sig}^j)N_{\sig}^{k)}
         + 2v_{\sig \mu}^{(i} 
           (\chi_{\sig}^j \sig^{\mu} \chib_{\sig}^{k)}) \right] \non
 && -\1{16}W_{ijkl}(C_{\sig})  
      \chi_{\sig}^{(i}\chi_{\sig}^j\chib_{\sig}^k\chib_{\sig}^{l)} . 
    \label{sig_Phi-V}
\eeq
The path integrals of $D_{\sig}^i$, 
$\lam_{\sig}^i$ and $\lamb_{\sig}^i$ 
give delta functions, $\delta(C_{\Phi i} - W_i(C_{\sig}))$, 
$\delta(\chi_{\Phi i} - W_{ij}(C_{\sig})\chi_{\sig}^j)$ and 
$\delta(\chib_{\Phi i} - W_{ij}(C_{\sig})\chib_{\sig}^j)$,  
respectively. 
The equation $C_{\Phi i} - W_i(C_{\sig})=0$ can be 
solved for $C_{\sig}$ uniquely: 
$C_{\sig}^i = \hat C_{\sig}^{i}(C_{\Phi})$, 
on the assumption that $\det {\bf W}(C_{\sig}) \neq 0$,
where the Hesse matrix ${\bf W}$ is defined by 
\beq
 ({\bf W})_{ij} (C_{\sig}) \defeq W_{ij} (C_{\sig})
          = {\del^2 W \over \del \sig^i \del \sig^j} (C_{\sig}).     
\eeq
The fermionic fields $\chi_\sig$ are given by 
\beq
 \hat\chi_{\sig}^{i} = W(\hat C_{\sig})^{ij} \chi_{\Phi j} , 
\eeq
where the inverse of the Hesse matrix is defined by 
\beq
  W^{ij} \defeq ({\bf W}^{-1})^{ij}.
\eeq
The path integrals over 
$C_{\sig}^i$, $\chi_{\sig}^i$ and $\chib_{\sig}^i$ 
can be performed trivially because of the delta functions.  
These integrals leave the factors   
$|\det {\bf W} (\hat C_{\sig})|^{-1}$, 
$(\det {\bf W} (\hat C_{\sig}))^2$ and 
$(\det {\bf W} (\hat C_{\sig}))^2$, respectively.
The remaining fields $M_{\sig}^i$, $N_{\sig}^i$ and 
$v_{\sig \mu}^i$ are quadratic: 
\beq
 && -\1{4} W_{ij} \left[
     (M_{\sig} - \hat{M}_{\sig})^i (M_{\sig} - \hat{M}_{\sig})^j 
   + (N_{\sig} - \hat N_{\sig})^i (N_{\sig} - \hat N_{\sig})^j 
     \right. \non
  && \left. \hspace{1cm} 
   - (v_{\sig} - \hat v_{\sig})^i \cdot (v_{\sig} - \hat v_{\sig})^j 
    \right] \non
 && + \1{4} W_{ij} \left[
  \hat M_{\sig}^{i} \hat M_{\sig}^{j} + \hat N_{\sig}^{i} \hat N_{\sig}^{j} 
  - \hat v_{\sig}^{i} \cdot \hat v_{\sig}^{j} \right]   , 
\eeq 
where we have defined 
\beq
 \hat M_{\sig}^{i} 
   &=& W^{ij} M_{\Phi j} 
     + \2{4} W^{ij} W_{jkl} W^{km} W^{ln} 
       (\chi_{\Phi m}\chi_{\Phi n} 
      - \chib_{\Phi m}\chib_{\Phi n}), \non 
 \hat N_{\sig}^{i} 
   &=& W^{ij} N_{\Phi j} 
     + \1{4} W^{ij} W_{jkl} W^{km} W^{ln}
       (\chi_{\Phi m}\chi_{\Phi n} 
      + \chib_{\Phi m}\chib_{\Phi n}), \non
 \hat v_{\sig \mu}^{i} 
   &=& W^{ij} v_{\Phi j \mu}
     -\1{2} W^{ij} W_{jkl} W^{km} W^{ln}
            (\chi_{\Phi (m} \sig_{\mu} \chib_{\Phi n)}) .\label{sig*}
\eeq
Here, we have used the notation $W_i = W_i(\hat C_{\sig})$, etc. 
The path integrals over $M_{\sig}^i$, $N_{\sig}^i$ and 
$v_{\sig \mu}^i$ provide 
the factors $(\det {\bf W})^{-\1{2}}$, $(\det {\bf W})^{-\1{2}}$ 
and $[(\det {\bf W})^{-\1{2}}]^4$, respectively. 
Thus, the result of the path integral can be written as 
$A \exp (i \int d^4x \Gamma)$, 
where $A$ and $\Gamma$ are as follows: 
\beq
 A &=& 
      {\det {\bf W} \over |\det {\bf W}|} 
    = {\rm sign} (\det {\bf W}) , \label{sign}\\
 \Gamma &=& 
  \1{2} {\hat C_{\sig}(C_{\Phi})}^i D_{\Phi i} 
  + \1{4}W^{ij}( M_{\Phi i}M_{\Phi j} 
               + N_{\Phi i}N_{\Phi j} 
               - v_{\Phi i}\cdot v_{\Phi j} ) \non 
 && - \1{2}W^{ij} (\chi_{\Phi (i} \lam_{\Phi j)} 
              + \chib_{\Phi (i} \lamb_{\Phi j)}) \non
 && + \1{8} W^{ij} W_{jkl} W^{km} W^{ln} 
   \left[i (\chi_{\Phi m}\chi_{\Phi n}
               -\chib_{\Phi m}\chib_{\Phi n}) M_{\Phi i}
         + (\chi_{\Phi m}\chi_{\Phi n} 
               + \chib_{\Phi m}\chib_{\Phi n})N_{\Phi i}
        \right. \non
     && \left. \hspace{3cm}
        + 2v_{\Phi i \mu} (\chi_{\Phi (m} 
             \sig^{\mu} \chib_{\Phi n)})\right] 
                   \non
 && +\1{16} 
    \left(3 W^{ij} W_{ikl}W_{jmn} - W_{klmn} \right)
    W^{ko}W^{lp}W^{mq}W^{nr}
    \chi_{\Phi (o}\chi_{\Phi p}\chib_{\Phi q}\chib_{\Phi r)} . 
 \label{explicit_form}
\eeq
When the correspondence between $C_{\sigma}$ and 
$C_{\Phi}$ is not unique, we should choose the branch 
where $\det {\bf W}(C_{\sig})$ has a definite sign. 

To prove the theorem, 
we calculate the D-term of $U(\Phi)$ 
by solving Eq.~(\ref{sigb}) explicitly,
and compare with $\Gamma$ in Eq.~(\ref{explicit_form}).
Using Eq.~(\ref{f(V)}), we can express Eq.~(\ref{sigb}) 
by components as
\beq
 0 &=& \Phi_i - W_i (\sig^1,\cdots,\sig^n) \non 
  &=&  C_{\Phi i} - W_i(C_{\sig}) 
   + i \th (\chi_{\Phi i} - W_{ij}  \chi_{\sig}^j) 
   - i \thb(\chib_{\Phi i} - W_{ij}  \chib_{\sig}^j) \non
 &&+ \2{2} \th\th 
  \left[ M_{\Phi i}+ iN_{\Phi i} 
       - W_{ij}(M_{\sig}^j+iN_{\sig}^j) 
       + \2{2}W_{ijk} \chi_{\sig}^j\chi_{\sig}^k \right] \non
 &&- \2{2} \thb\thb 
  \left[ M_{\Phi i}- iN_{\Phi i} 
       - W_{ij} (M_{\sig}^j -iN_{\sig}^j) 
       - \2{2}W_{ijk} \chib_{\sig}^j\chib_{\sig}^k \right] \non
 &&- \th \sig^{\mu} \thb 
  \left[v_{\Phi i \mu} - W_{ij} v_{\sig \mu}^j 
     - \1{2}W_{ijk}
         (\chi_{\sig}^{(j} \sigma_{\mu} 
          \chib_{\sig}^{k)})\right]\non
 &&+ i\th\th\thb
   \left[ \lamb_{\Phi i} - W_{ij} \lamb_{\sig}^j 
      + \2{2}W_{ijk} \chib_{\sig}^{(j} (M_{\sig}+iN_{\sig})^{k)}
      +\1{2}W_{ijk} \sigb^{\mu} 
            \chi_{\sig}^{(j} v_{\sig \mu}^{k)} 
      +\1{6}W_{ijkl} \chib_{\sig}^{(j}\chi_{\sig}^k\chi_{\sig}^{l)} 
   \right]\non
 &&- i\thb\thb\th
   \left[ \lam_{\Phi i} - W_{ij} \lam_{\sig}^j 
      - \2{2}W_{ijk} \chi_{\sig}^{(j} (M_{\sig}-iN_{\sig})^{k)}
      -\1{2}W_{ijk} \sig^{\mu} \chib_{\sig}^{(j} v_{\sig \mu}^{k)} 
      +\1{6}W_{ijkl} \chi_{\sig}^{(j}\chib_{\sig}^k
                     \chib_{\sig}^{l)}
   \right]\non
 &&+ \1{2} \th\th\thb\thb
    \bigg[ D_{\Phi i} - W_{ij}(C) D_{\sig}^j 
         + W_{ijk}(C)(\chi_{\sig}^{(j}\lam_{\sig}^{k)} 
                  + \chib_{\sig}^{(j}\lamb_{\sig}^{k)}) 
    \bigg.\non
  &&  \bigg. \hspace{1cm}
       - \1{2} W_{ijk}(C) 
      (M_{\sig}^j M_{\sig}^k + N_{\sig}^j N_{\sig}^k 
     - v_{\sig}^j\cdot v_{\sig}^k ) 
          \bigg.\non
  &&\hspace{1cm} \bigg.
    - \1{4} W_{ijkl}(C) 
       \{ i(\chi_{\sig}^{(j}\chi_{\sig}^k 
          - \chib_{\sig}^{(j}\chib_{\sig}^k) M_{\sig}^{l)} 
        +  (\chi_{\sig}^{(j}\chi_{\sig}^k 
        + \chib_{\sig}^{(j}\chib_{\sig}^k) N_{\sig}^{l)}
        +  2 {v_{\sig \mu}^{(j}}
           (\chi_{\sig}^k\sig^{\mu}\chib_{\sig}^{l)}) \} 
   \bigg.\non
  &&\hspace{1cm} \bigg.
   - \1{8}W_{ijklm}(C) 
     \chi_{\sig}^{(j}\chi_{\sig}^k\chib_{\sig}^l\chib_{\sig}^{m)} 
                      \bigg].
\eeq
These equations can be solved by components as
\beq
 && C_{\Phi i} = W_i ( \hat C_{\sig}), \non 
 &&  \hat \chi_{\sig}^i = W^{ij} \chi_{\Phi j},\non
 && \hat M_{\sig}^i 
   = W^{ij} M_{\Phi j} 
     + \2{4} W^{ij} W_{jkl} W^{km} W^{ln} 
       (\chi_{\Phi m}\chi_{\Phi n} 
      - \chib_{\Phi m}\chib_{\Phi n}), \non 
 &&\hat N_{\sig}^i
   = W^{ij} N_{\Phi j} 
     + \1{4} W^{ij} W_{jkl} W^{km} W^{ln}
       (\chi_{\Phi m}\chi_{\Phi n} 
      + \chib_{\Phi m}\chib_{\Phi n}), \non
 &&\hat v_{\sig \mu}^i 
   = W^{ij} v_{\Phi j \mu}
     -\1{2} W^{ij} W_{jkl} W^{km} W^{ln}
            (\chi_{\Phi (m} \sig_{\mu} \chib_{\Phi n)}) .
    \label{sol.of_sigma}
\eeq
Here we have not written the solutions for 
$\hat \lam_{\sig}$ and $\hat D_{\sig}$, 
since they disappear when the first and second equations are 
substituted into Eq.~(\ref{sig_Phi-V}). 
The first equation can be solved implicitly by $\hat C_{\sig}$. 
By substituting these equations into Eq.~(\ref{sig_Phi-V}), 
we confirm that it coincides with Eq.~(\ref{explicit_form}).  
(Q.E.D.)

We give the chiral superfield version of Theorem 1 
in Appendix A. 

\bigskip
We can also prove the matrix version of Theorem 1,
which will be applied to integration over 
the non-Abelian vector superfields in the next section.

{\bf Corollary 1.}
(The matrix version of the quantum Legendre transformation.)\\
Let $\Sig$ and $\Psi$ be matrix-valued vector superfields 
and let $W(\Sig)$ be a scalar function of $\Sigma$.\footnote{
For example, it is a single trace of a function $w$, 
$W(\Sig) = \tr w(\Sig)$, but it need not be a single trace.} 
Then, 
\beq
  \int [d\Sigma] \exp \left[i \int d^4 x d^4 \th  
       \left( \tr (\Sig\Psi) - W(\Sig) \right) \right] 
 = \exp\left[i \int d^4x d^4\th  \;U(\Psi) \right] .
\eeq
Here $U(\Psi)$ is defined as 
\beq
 U(\Psi) = \tr (\hat{\Sig}(\Psi) \Psi) 
         - W\left(\hat{\Sig}(\Psi)\right) ,
\eeq
where $\hat{\Sig}$ is a solution of the stationary equation 
\beq
 {\del \over \del \Sig_{ij}} 
  \bigg[\tr (\Sig \Psi - W(\Sig))\bigg]|_{\Sig =\hat \Sig} 
 = \Psi_{ij} - \left( \del_{\Sig}W(\hat \Sig) \right)_{ij} = 0 .
\eeq
(Proof) 
The proof is trivial, if we rewrite 
$\tr (\Sigma \Psi) = \Sigma_{ij} \Psi_{ji}$ and 
identify a pair of indices ${ij}$ with an index 
in Theorem 1. (Q.E.D.)

If the scalar function $W(\Sig)$ 
is written in a single trace as 
$W(\Sig) = \tr w(\Sig)$ with a function $w$,  
its derivative with respect to $\Sigma$ can be written as 
$\del_{\Sig_{ij}}W(\Sig) = {w\pri(\Sig)^T}_{ij}$.
Note that, as in Eq.~(\ref{sign}), 
we must assume that the Hessian has 
a definite sign. 
However, this is automatically the case 
for the example in the next section.

\bigskip
We derive one more useful corollary of Theorem 1, 
which will be used in the proof of Theorem 2 in the 
next section.
 
{\bf Corollary 2.}\\
Let $\sig^i$ ($i=1,\cdots,n$) 
be vector superfields and $W$ be a function of $\sig^i$. 
Then,  
\beq
 \int [d\sig] \exp\left[i \int d^4x d^4\th  
    \;W(\sig^1,\cdots,\sig^n)\right]
 = 1 
\eeq
when the Hessian 
${\bf W}_{ij} = \del_i\del_j W(C_{\sig})$ 
is positive definite.\\ 
(Proof) 
By Theorem 1, 
the path integral of $\sig$ is given by 
the solution $\hat\sig^i$ of the stationary equation, 
$W_i(\sig^1,\cdots,\sig^n)|_{\sig^i = \hat\sig^i}= 0$.  
By using Eq.~(\ref{f(V)}), 
it can be written in components as 
\beq
 W_i (\hat C_{\sig}^1, \cdots, \hat C_{\sig}^n) = 0 ,\hspace{1cm}
 \hat \chi_{\sig}^i = \hat M_{\sig}^i = \hat N_{\sig}^i 
  = \hat v_{\sig}^i = \hat \lambda_{\sig}^i 
  = \hat D_{\sig}^i = 0 ,
\eeq
if the Hessian, $\det {\bf W}(\hat C_{\sig})$, 
is not zero. The solutions of the first equations 
$\hat \sig^i = \hat C_{\sig}^i$
are c-numbers. Therefore, the left-hand side is
\beq
 ({\rm LHS}) &=& 
   \exp\left[i \int d^4x d^4\th  
    \;W(\hat\sig^1,\cdots,\hat\sig^n)\right] \non 
 &=& \exp\left[i \int d^4x d^4\th  
    \;W(\hat C_{\sig}^1,\cdots,\hat C_{\sig}^n)\right] 
 = 1 
\eeq
when the Hessian is positive definite. (Q.E.D.)\\

\bigskip

In the remainder of this section, we prove that the path 
integral measure is invariant under a change of variables 
in supersymmetric theories.

{\bf Lemma 1.} (The invariance of the measure.)\\
When two sets of $n$ vector superfields are 
related by a nonsingular local transformation, 
\beq
 \sigma^i = f^i(V^1,\cdots,V^n), \;\;\;\; (i=1,\cdots,n)
\eeq 
the measure is invariant: 
\beq
 [d \sigma] = [d V] . \hspace{1cm} 
\eeq
(Proof) 
The measures of the superfields are defined by 
their component fields as 
\beq
 &&[d\sig] \defeq [\prod_i d\sig^i],\;\; 
   d\sig^i \defeq 
   dC_{\sig}^i d\chi_{\sig}^i d\chib_{\sig}^i dM_{\sig}^i dN_{\sig}^i
    d v_{\sig m}^i d\lamb_{\sig}^id\lam_{\sig}^i dD_{\sig}^i,\;\; \non
 &&[dV] \defeq [\prod_i dV^i],\;\; 
   dV^i \defeq  
    dC_V^i d\chi_V^i d\chib_V^i 
    dM_V^i dN_V^i
    d V_{V m}^i d\lamb_V^i 
    d\lam_V^i dD_V^i.\;\; \label{measure}
\eeq
For convenience, we reorder them as 
$[d\sig]=[\prod_i dC_{\sig}^i\prod_i d\chi_{\sig}^i \cdots]$, 
etc. From Eq.~(\ref{f(V)}), the superfield equation, 
$\sig^i = f^i(V)$, can be written in component fields as\footnote{
Note that the ranges of the variables remain unchanged, 
except for $C_{\sig}$ and $C_V$, 
provided that ${f^i}_j(C_V)\neq 0$.}   
\beq 
 &&C_{\sig}^i = f^i(C_V), \;\;
   \chi_{\sig}^i = {f^i}_j(C_V) \chi_V^j,\;\; 
   \chib_{\sig}^i = {f^i}_j(C_V) \chib_V^j,\non 
 &&M_{\sig}^i = {f^i}_j(C_V) M_V^j 
   - \2{4} {f^i}_{jk}(C_V) 
           (\chi_V^j \chi_V^k 
          - \chib_V^j\chib_V^k),\non 
 &&N_{\sig}^i = {f^i}_j(C_V) N_V^j
   - \1{4} {f^i}_{jk}(C_V) 
           (\chi_V^j \chi_V^k 
           + \chib_V^j\chib_V^k),\non
 &&v_{\sig \mu}^i = {f^i}_j(C_V) v_{V \mu}^j + 
  \1{2} {f^i}_{jk}(C_V) 
    (\chi_V^{(j} \sig_{\mu} \chib_V^{k)}),
     \;\;\cdots.
\eeq
The measures in the two set of coordinates are related by     
\beq
 [d\sig] = J [d V],
\eeq
where $J$ is the Jacobian, $J={\rm sdet}\, M$. 
Here, sdet is a superdeterminant of a $16 n \times 16 n$ 
supermatrix $M$ 
($8$ bosonic and $8$ fermionic components 
for each vector superfield),  
whose diagonal blocks are all ${f^i}_j(C_V)$, 
and whose off-diagonal part is nilpotent: 
\beq 
 M = {\bf 1}_{16}\otimes {\bf f}(C_V) + M\pri,
\eeq  
where ${({\bf f})^i}_j = {f^i}_j$ 
and $M\pri$ is a nilpotent matrix. 
Then, $J$ can be written as  
\beq
 J = {\rm sdet}\, M = \exp({\rm str} \log M), 
\eeq
where ${\rm str}$ is a supertrace. 
This can be calculated as\footnote{
If we write $M = A B$, 
then from the equation 
$\log M = \log (A B) 
     = \log A + \log B + \1{2} [\log A,\log B] + \cdots$, 
we obtain the formula  
${\rm sdet}\, M = {\rm sdet}\, A \cdot {\rm sdet}\, B$.
}
\beq
 {\rm sdet} M &=& {\rm sdet}\, A \cdot {\rm sdet}\, B, \non
 \log A &=& \log ({\bf 1}_{16} \otimes {\bf f}(C_V)), \non
 \log B &=& \log \left({\bf 1}_{16n} 
            + ({\bf 1}_{16} \otimes {\bf f}^{-1})M\pri\right) 
         = \sum_{m>0} {(-1)^m \over m} 
           \left( ({\bf 1}_{16} \otimes {\bf f}^{-1})M\pri\right)^m ,\;\;
\eeq
where ${\bf f}^{-1}$ is the inverse matrix of ${\bf f}$. 
Since the supertrace of $\log A$ is zero 
due to the cancellation of bosonic and fermionic 
contributions ($8-8=0$), ${\rm sdet} A = 1$.   
Because $M\pri$ is nilpotent, $\log B$ is also nilpotent, 
and hence ${\rm sdet}\, B = 1$ and  $J=1$. 
(Q.E.D.) 

The invariance of the measure also holds for matrix-valued 
superfields. 

{\bf Lemma 2.}  
(The invariance of the measure for matrix-valued superfields.)\\
Let $\Sigma$ and $V$ be matrix valued vector superfields 
related by a local transformation $\Sigma = f(V)$. Then
\beq
 [d \Sigma] = [d V], \hspace{1cm} 
\eeq
where the measures are defined by 
$[d \Sigma] = \prod_{ij} d \Sigma_{ij}$ and 
$[d V] = \prod_{ij} d V_{ij}$.\\
(Proof) The proof is trivial, 
since $\Sigma_{ij} = f^{ij}(V_{11},\cdots,V_{NN})$, 
where $f^{ij}$ are functions of $V_{11},\cdots,V_{NN}$. 
(Q.E.D.)

\section{Applications to Nonlinear Sigma Models}
In this section, we apply our results to show the 
equivalence of two lagrangians with 
and without auxiliary fields. 
Firstly, we apply the theorems obtained in the 
last section to non-Abelian gauge theories, 
where the vector superfields are auxiliary fields. 
Consider the $N\times M$ matrix valued chiral superfields 
$\Phi$ belonging to the fundamental representation of 
global symmetry $U(N)$ and the anti-fundamental 
representation of the gauge group $U(M)$.
The action of the global and gauge symmetries are as follows:
\beq
 \Phi \to g_L \Phi {g_R}^{-1},
\eeq
where $g_L$ is an $N\times N$ matrix of $U(N)$ and 
$g_R$ is an $M\times M$ matrix of $U(M)$.
The gauge invariant lagrangian of $\Phi$ 
interacting with the vector superfield 
$V = V^A T_A$,
where $T_A$ is the Lie algebra of $U(M)$, reads
\beq
 {\cal L} 
 =\int d^4 \th K_0 (\Ph,\Ph\dagg,V)
 = \int d^4 \th  
      \left( \tr (\Phi\dagg\Phi e^V) 
            - c\,\tr V \right).
\eeq
Here $c$ is the Fayet-Iliopoulos (FI) parameter.

As described in the Introduction, 
we can obtain the K\"{a}hler potential of the nonlinear 
sigma model when we eliminate $V$ with its equation of 
motion. As an application of the theorem obtained 
in the last section, we show that this procedure is 
also justified at the quantum level.  
We obtain the next corollary from Corollary 1. 

{\bf Corollary 3.} 
(Integrating out the non-Abelian vector superfields.)\\
Let $\Phi$ be an $N\times M$ matrix-valued chiral superfield 
and $V$ be a matrix-valued vector superfield $V = V^A T_A$, 
where $T_A$ is the $M\times M$ matrix of a Lie algebra.
Then 
\beq
  \int [d V] \exp \left[i \int d^4 x d^4 \th  
      \left( \tr (\Phi\dagg\Phi e^V) 
            - c\,\tr V \right) \right] 
 = \exp \left[ i \int d^4 x d^4 \th
                   \,c \, \tr \log (\Phi\dagg\Phi) \right] .
\eeq
(Proof) 
In Corollary 1, we replace $\Sig$ and ${\Psi}$ 
by $e^V$ and $\Phi\dagg \Phi$, respectively.
Then the left-hand side can be calculated as 
\beq
 ({\rm LHS})
 &=& \int [d \Sig] \exp \left[ i \int d^4 x d^4 \th  
      \left(\tr ( \Sig \Psi) - c \,\tr \log \Sig \right)
                \right]  \non
 &=& \exp \left[ i \int d^4 x d^4 \th \; U(\Psi) \right] 
 = ({\rm RHS}), 
\eeq
where $U$ has been defined as 
$U(\Psi) = \tr (\hat \Sig(\Psi)\Psi) 
- c \,\tr \log \hat\Sig(\Psi)$ 
and $\hat \Sig(\Psi) = c \,\Psi^{-1}$. 
We have used Lemma 2 in the first equality and 
Corollary 1 in the second equality. 
The constant term disappears under integration over $\th$: 
$\int d^4 \th ( iNc - iNc \log c) = 0$. 
It is proven in Appendix B that 
the Hessian $\det {\bf W}$ in Eq.~(\ref{sign}) has 
a definite sign.
(Q.E.D.)

\bigskip
Secondly, we apply the theorems to 
show the uniqueness of the effective lagrangian 
of the supersymmetric nonlinear sigma models 
on compact K\"{a}hler manifolds. 

When a global symmetry $G$ is spontaneously broken, 
there appear Nambu-Goldstone (NG) bosons 
corresponding to the broken generators. 
Since a chiral superfield has two real scalar fields, 
the partner of a NG-boson may not be a NG-boson. 
If this is the case, 
the partner of the NG-boson is called a 
quasi-Nambu-Goldstone (QNG) boson~\cite{BPY}. 
When there is no gauge symmetry, it is known that there must be 
at least one QNG boson~\cite{noncompact}, 
and the K\"{a}hler potential of 
the effective lagrangian can be written as an arbitrary 
function of the $G$ invariants~\cite{BPY,BKMU,noncompact}. 
QNG bosons correspond to the {\it non-compact} 
directions of the target space and introduce 
an arbitrariness into the K\"{a}hler potential.
On the other hand, the K\"{a}hler potential 
has no arbitrariness~\cite{BKMU,IKK} 
if the target space is {\it compact}. 
Our task in the remainder of this section is 
to show that the arbitrariness disappears 
in the process of eliminating the non-compact directions.
We introduce gauge fields to eliminate the QNG-bosons 
corresponding to non-compact directions. 
When matrix-valued chiral superfields $\Phi$ 
acquire the vacuum expectation value, 
the most general effective K\"{a}hler potential 
can be written as~\cite{BKMU,noncompact} 
\beq
 K_0 (\Phi, \Phi\dagg) 
 = f \left( \tr (\Phi\dagg\Phi) \right) , \label{kahler1} 
\eeq
where $f$ is an {\it arbitrary} function introduced  
by the QNG bosons.\footnote{  
The low-energy interactions of NG and QNG bosons were 
obtained as low energy theorems~\cite{HNOO}.
Low-energy scattering amplitudes of NG bosons 
coincide with those of non-supersymmetric theories, 
despite the existence of the arbitrariness. 
Those of QNG bosons coincide with those of corresponding NG bosons.
Moreover, the arbitrariness appears in 
the interactions of NG and QNG bosons.
}
When we introduce gauge fields to eliminate 
non-compact directions, the gauge invariant 
K\"{a}hler potential reads
\beq
 K_0 = f \left( \tr (\Phi\dagg\Phi e^V) \right) 
 - c\, \tr V.  \label{kahler2}
\eeq 
The arbitrariness of the K\"{a}hler potential comes from 
the existence of the QNG bosons. 
Hence, if the gauge fields absorb all the QNG bosons, 
the target manifold becomes compact~\cite{IKK}, 
and the arbitrariness disappears after integrating 
out the vector superfield. We show that
this actually is the case.\footnote{
It is known that, at the classical level, 
this arbitrariness must disappear when $V$ is 
eliminated by using its equation of motion~\cite{BKY}.}  

Firstly, we generalize Corollary 1 to the case in which 
an arbitrary function appears.

{\bf Theorem 2.} \\
Let $\Sigma$ and $\Psi$ be $N\times N$ 
matrix-valued vector superfields.
Then the arbitrary function $f$ disappears for 
$W(\Sig) = c \,\tr \log \Sig \defeq c \,\tr w(\Sig)$:
\beq
 &&\int [d\Sig] \exp \left[i \int d^4 x d^4 \th  
        \bigg( f(\tr (\Sig \Psi)) - W(\Sig) \bigg) \right] \non
 &&= \int [d\Sig] \exp \left[i \int d^4 x d^4 \th  
           \left( \tr (\Sig \Psi) - W(\Sig) \right) \right] . 
\eeq
(Proof)
Let us introduce the Legendre transform $\tilde W$ 
of the arbitrary function $f$: 
\begin{equation}
 f(\phi) = \rho\phi - \tilde W(\rho), \qquad 
 \phi - \tilde W\pri(\rho) = 0 .
\end{equation}
Then, the arbitrary function of $\tr (\Sig \Psi)$ can be 
linearized as 
\beq
&&\exp \left[i \int d^4 x d^4 \th f(\tr (\Sig \Psi)) \right] \non
&&=\int [d\rho] \exp \left[i \int d^4 x d^4 \th
    ( \rho\, \tr (\Sig \Psi) - \tilde W(\rho) )
  \right]. \nonumber 
\eeq
By substituting this expression into the left-hand side 
and introducing the rescaled vector superfields 
$\Sig\pri =\rho\Sig$, we obtain\footnote{
Note that the Jacobian is trivial by Lemma 1
}
\beq
 ({\rm LHS})
 &=& \int [d\Sig d\rho] \exp \left[i \int d^4 x d^4 \th
    ( \rho\, \tr (\Sig \Psi) - \tilde W(\rho) - c\,\tr w(\Sig))
  \right] \non 
 &=& \int [d\Sig\pri d\rho] \exp \left[i \int d^4 x d^4 \th
    \left(\tr (\Sig\pri \Psi) - \tilde W (\rho) 
          - c\,\tr w \left(\rho^{-1}\Sig\pri\right) 
    \right) \right] \non
 &=& \int [d\Sig\pri] 
   \exp \left[i \int d^4 x d^4 \th
     \left(\tr (\Sig\pri \Psi) - c\,\tr w(\Sig\pri)\right) \right] 
 = ({\rm RHS}) ,
      \label{th3_proof} 
\eeq
where in the third equality, 
the decoupled $\rho$ integral is evaluated by 
using Corollary 2: 
\beq
 \int [d\rho]\exp \left[ -i \int d^4 x d^4 \th 
       \left(\tilde W(\rho) 
            - c N w(\rho)\right) \right] =1. 
\eeq
(Q.E.D.) 

Note that the relation $W(\Sig) = c \log \Sig$ is essential 
to decouple the $\rho$ integration, 
and this theorem can not be generalized to 
the case of arbitrary $W$.
However, it is sufficient for our purpose 
to show the following corollary.

{\bf Corollary 4.} (The uniqueness of the seed metric.)\\
Consider the situation considered in Corollary 3. Then, 
\beq
 &&\int [d V] \exp \left[i \int d^4 x d^4 \th  
    \left( f \left(\tr (\Phi\dagg\Phi e^V) \right)  
                - c \,\tr V \right) \right] \non
 &=& \int [d V] \exp \left[i \int d^4 x d^4 \th  
           ( \tr (\Phi\dagg\Phi e^V) - c \,\tr V ) 
         \right] . 
\eeq
(Proof) This result directly follows from Theorem 2 
with the identification of $\Sig = e^V$ and 
$\Psi =\Phi\dagg\Phi$. The triviality of the 
Jacobian follows from Lemma 2. (Q.E.D.) 

This corollary shows that the arbitrariness 
in effective theories without a gauge interaction 
like Eq.~(\ref{kahler1}), 
disappears through the introduction of vector superfields like  
Eq.~(\ref{kahler2}). 
Hence, we obtain the same K\"{a}hler potentials 
by Corollary 3.
Since the local gauge symmetry eliminates 
the degrees of freedom in non-compact directions,
the obtained manifold becomes compact.

\bigskip
Before closing this section, 
we give a simple example: 
$SO(N)/SO(N-2)\times U(1)$~\cite{HN}. 
Consider chiral superfields $\vec{\phi}$ belonging 
to the vector representation of $SO(N)$. 
We impose the 
$O(N)$ invariant constraint $\vec{\phi}^2=0$ to define a 
manifold. This manifold is non-compact, since the defining 
condition is invariant under the scale transformation
$\vec{\phi}\rightarrow \lambda\vec{\phi}$. We can 
eliminate this scale invariance by introducing 
a $U(1)$ gauge field for the local scale 
(complex phase) symmetry. 
The gauged lagrangian is
\beq
 {\cal L} =  \int d^4 \th 
 \left(f(e^V \vec{\phi}^{\,\dagger}\vec{\phi}) 
       - c V \right)
 + \left(\int d^2 \th \,
    g \phi_0 \vec{\phi}^{\,2} + {\rm c.c.} \right), 
\eeq
where $V$ and $\phi_0$ are auxiliary vector 
and chiral superfields. 
We use a rather unconventional $SO(N)$ basis 
such that the invariant metric becomes
\beq
 J = \pmatrix{0 & {\bf 0}      & 1 \cr
         {\bf 0}& {\bf 1}_{N-2}&{\bf 0} \cr
              1 & {\bf 0}      & 0 } . \label{J-SO} 
\eeq
Then the $\phi_0$ integration gives the F-term constraint
\beq
 \vec{\phi}^{\,2} =\vec{\phi}^T J \vec{\phi}
 = 2 a b + \ph^2 = 0, \label{F-con.}
\eeq
where we have put $\vec{\phi}^{\,T} = (a,\ph^i,b)$.  
This equation can be easily solved as 
$\vec{\phi}^{\,T} = (a,\ph^i,-\ph^2/2a)$. 
Our theory has a gauge invariance which has the degree 
of freedom of a chiral superfield. To fix the gauge, 
we set $a=1$, giving  
$\vec{\phi}^{\,T} = (1,\ph^i,-\ph^2/2)$.
By Corollaries 4 and 3, 
the integration of $V$ gives the K\"{a}hler potential, 
\beq
 K= c \log(\vec{\phi}^{\,\dagger}\vec{\phi}) 
  = c \log \left(1 + |\ph|^2 + \1{4}\ph^{\dagger2} \ph^2 \right).
\eeq
This is the K\"{a}hler potential of 
a compact homogeneous K\"{a}hler manifold, 
$Q_{N-2}({\bf C}) = SO(N)/SO(N-1)\times U(1)$~\cite{IKK,Ni2}, 
which is called the ``quadratic surface''~\cite{KN}.
If we drop the F-term constraint, 
this becomes the K\"{a}hler potential 
for the Fubini-Study metric of ${\bf C}P^{N-1}$.
Hence, $Q_{N-2}({\bf C})$ is holomorphicaly embedded into 
${\bf C}P^{N-1}$ by the F-term constraint Eq.~(\ref{F-con.}).

Generalization to a broader class, 
namely the hermitian symmetric spaces, 
is discussed in a separate paper~\cite{HN}.

\section*{Acknowledgements}
We would like to thank Motonobu Torii and 
Tetsuji Kimura for comments.
The work of M.~N. is supported in part 
by JSPS Research Fellowships.

\begin{appendix}

\section{Chiral Superfield Version of Theorem 1}
In this appendix, 
we give the chiral superfield version of Theorem 1 
without any proof. 

{\bf Theorem 3.} (Chiral superfield version.)

Let $\ph^i$ and $\phi_i$ ($i=1,\cdots,n$) 
be chiral superfields and $W$ be a function of $\ph^i$. 
Then,  
\beq
&& \int [d\ph] \exp i \left[\int d^4x d^2\th  
    \;(\ph^i \phi_i - W(\ph^1,\cdots,\ph^n)) 
   +({\rm c.c.})\right] \non
&& = \exp i \left[\int d^4x d^2\th  \;U(\phi) + ({\rm c.c.})\right] .
\eeq
Here $U(\phi)$ is defined as 
\beq
 U(\phi) = \hat{\ph}^{i} (\phi) \phi_i 
  - W(\hat{\ph}^{1} (\phi),\cdots,\hat{\ph}^{n} (\phi)) ,
\eeq
where $\hat{\ph}$ is a solution of the stationary equation 
\beq
 {\del \over \del \ph^i} 
  (\ph^j \phi_j - W (\ph^1,\cdots,\ph^n)) |_{\ph^i =\hat\ph^{i}} 
 = \phi_i - W_i (\hat\ph^{1},\cdots,\hat\ph^{n}) = 0 .
\eeq

\section{Definiteness of the Sign of the Hessian}
Let $\Sig$ be an $N$ by $N$ matrix.  
In this appendix, for the case $W(\Sig) = c\, \tr \log \Sig$, 
we calculate the Hessian and show that its sign is definite.
The Hesse matrix consists of 
derivatives of $W$ with respect to two $\Sig_{ij}$ variables: 
\beq
 {\bf W}_{ij,kl}(C_{\Sig}) 
 = {\del^2 W (C_{\Sig})\over \del\Sig_{ij}\del\Sig_{kl}}
 = c {\del {\Sigma^{-1}}_{ji} \over \del \Sigma_{kl}} .
\eeq
By differentiating an identity 
$\delta_{ij} = \Sigma_{ik}{\Sigma^{-1}}_{kj}$, 
we obtain the formula
\beq
 {\del {\Sigma^{-1}}_{ji} \over \del \Sigma_{kl}} 
 = - {\Sigma^{-1}}_{jk} {\Sigma^{-1}}_{li} .
\eeq 
Then the Hesse matrix can be calculated as
\beq
 {\bf W}_{ij,kl}(C_{\Sig})
 = -c \Sig^{-1}_{jk}\Sig^{-1}_{li}. 
\eeq
Hence we can calculate the Hessian as\footnote{
When a matrix $M$ can be written as a tensor product of 
two $N$ by $N$ matrices $u$ and $v$ as $M = u \otimes v$,   
its determinant is $\det M = (\det u)^{N^2} (\det v)^{N^2}$.
} 
\beq
 \det_{N^2 \times N^2} {\bf W} 
 = (-c)^{N^2} (\det \Sig^{-1})^{2N^2}.  
\eeq
The sign of this is definite, 
as used in the proof of Corollary 3.

\end{appendix}


\end{document}